\documentclass{PoS}

\title{Physics with Charmonium - A few recent highlights of BESIII}

\ShortTitle{Physics with Charmonium at BESIII}

\author{\speaker{Johan Messchendorp} for the BESIII Collaboration\\
        KVI, University of Groningen, Zernikelaan 25, 9747 AA Groningen, The Netherlands.\\
        E-mail: \email{messchendorp@kvi.nl}}

\abstract{Despite the successes of the standard model, the non-perturbative dynamics of the strong interaction are not fully understood yet. 
Charmonium spectroscopy serves as an ideal tool to shed light on the dynamics of the strong interaction such as quark confinement and the generation of hadron masses. 
The BESIII collaboration studies extensively the strong interaction and various aspects that could shed light on physics beyond the standard model via 
copious e$^+$e$^-$ collisions at the BESIII/BEPCII facility in Beijing, China, in the charmonium mass regime. I present a few of the recent results with 
the emphasis on charmonium spectroscopy and decay studies using 106$\times$10$^6$ $\psi(3686)$ events.}

\FullConference{International Winter Meeting on Nuclear Physics,\\
		21-25 January 2013\\
		Bormio, Italy }

\begin{document}

\section{Introduction}

\subsection{The strong force and QCD}

\begin{figure} \includegraphics[width=\textwidth]{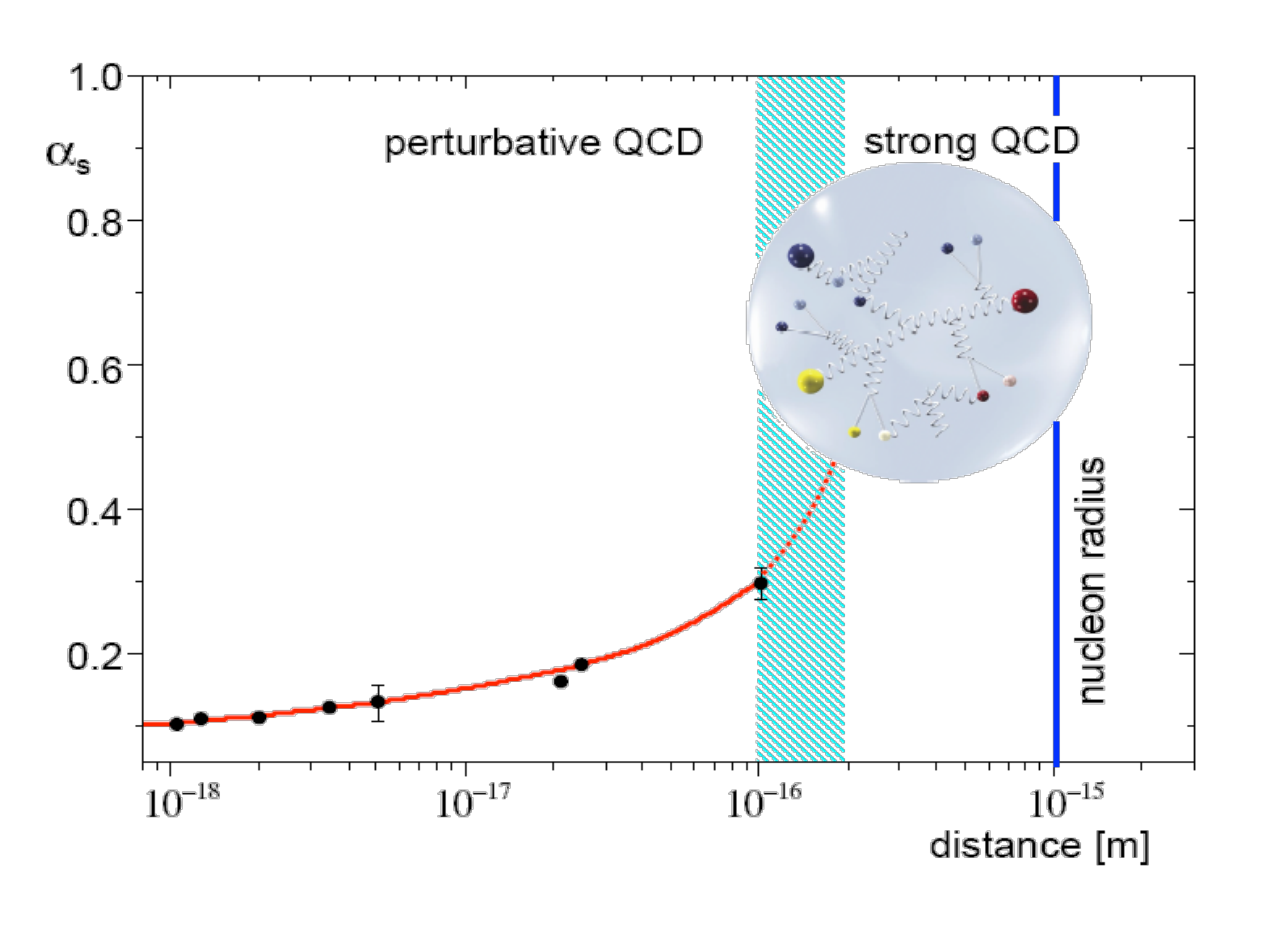}
\vspace*{-1.5cm}
\caption{The strong coupling constant, $\alpha_S$, as a function of the distance scale. 
Towards larger distances, QCD becomes non-perturbative, and gives rise to spectacular 
phenomena such as the generation of hadron masses and quark confinement.}
\label{fig_strong_coupling}
\end{figure}

The fundamental building blocks of Quantum Chromodynamics (QCD) 
are the quarks which interact with each other by exchanging gluons. QCD is well 
understood at short-distance scales, much shorter than the 
size of a nucleon ($<$~10$^{-15}$~m). In this regime, the basic 
quark-gluon interaction is sufficiently weak. In fact, many processes 
at high energies can quantitatively be described by perturbative QCD.
Perturbation theory fails when the distance among quarks becomes 
comparable to the size of the nucleon. Under these conditions, in the 
regime of non-perturbative strong QCD, the force among the quarks 
becomes so strong that they cannot be further separated (see illustration in Fig.~\ref{fig_strong_coupling}). As a 
consequence of the strong coupling, we observe the relatively heavy 
mass of hadrons, such as protons and neutrons, which is two orders of 
magnitude larger than the sum of the masses of the individual quarks. 
This quantitatively yet-unexplained behavior is related to the 
self-interaction of gluons leading to the formation of gluonic 
flux tubes connecting the quarks. As a consequence, quarks have 
never been observed as free particles and are confined within hadrons, 
i.e. the baryons containing three valence quarks or mesons containing 
a quark-antiquark pair. 

Besides the presence of conventional hadrons consisting of three quarks, baryons,
or two quarks, mesons, the intriguing color nature of QCD allows for the 
existence of gluon-rich matter, such as hybrids and glueballs, states composed
of more than three quarks, or weakly bound states composed of mesons, the so-called
molecular states. Although, experimental hints for their existence are available,
there is yet no unambiguous discovery of such states. A systematic approach 
exploiting a clean environment and high statistics data has a good discovery 
potential in the search for the new form of hadronic matter. 

\subsection{Charmonium: the positronium of QCD}

\begin{figure} \includegraphics[width=\textwidth]{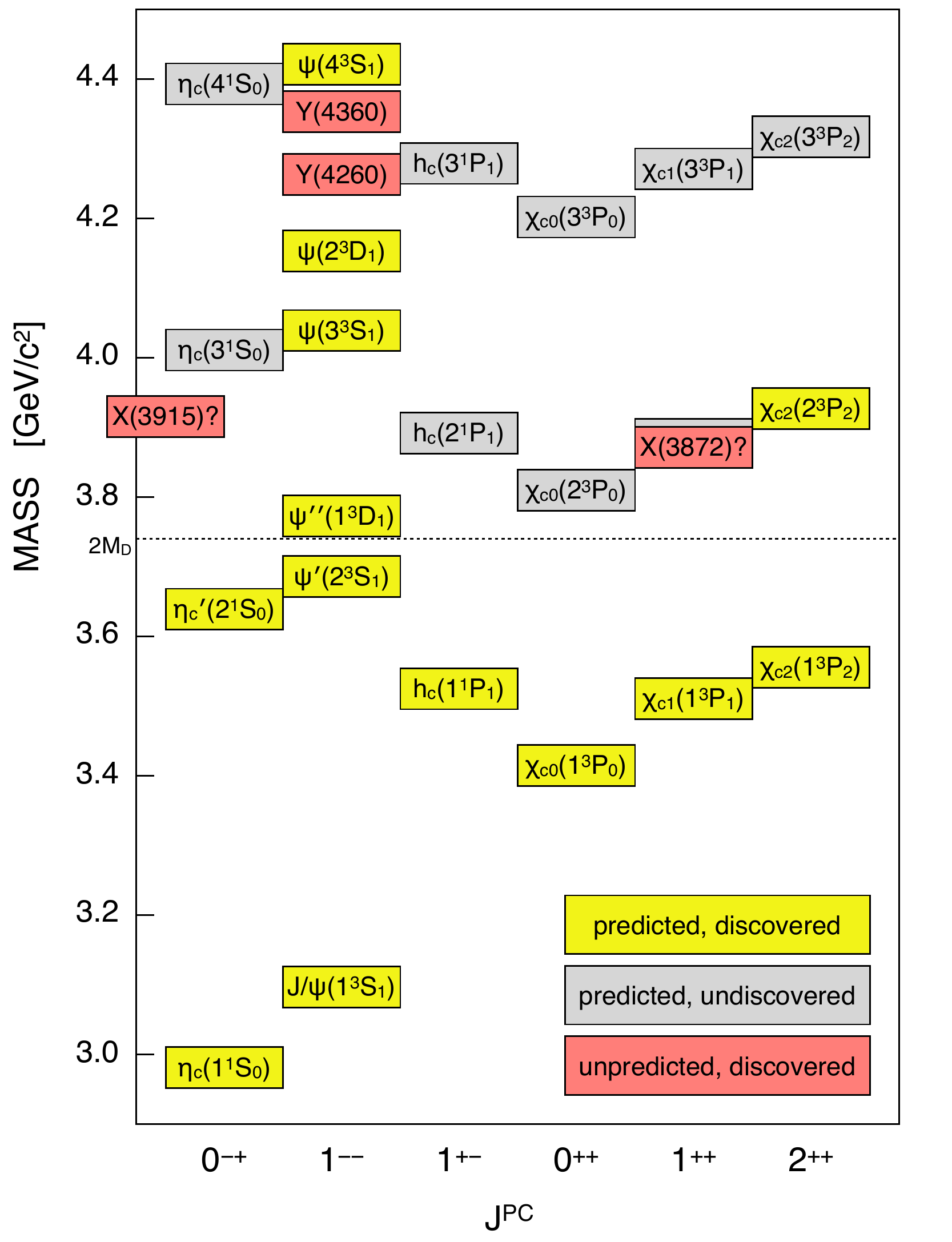}
\vspace*{-1cm}
\caption{The mass spectrum of charmonium(-like) states in the energy interval available in BESIII
and as a function of their spin-parity, $J^{PC}$. The yellow boxes represent charmonium states 
predicted by theory and confirmed by experiment. 
The grey boxes are those charmonium states that are predicted but not yet discovered. 
The red boxes are discovered charmonium-like states which nature is still mysterious.
The dashed line indicates the open-charm ($D\bar{D}$) threshold. The figure is taken
from a presentation by R.~Mitchell.}
\label{charmonium_spectrum}
\end{figure}

The BESIII collaboration exploits the bound state of a charm quark ($c$)  
and a charm antiquark ($\bar c$), known as charmonium, to study as part of an extensive
physics program the dynamics of the strong force in an energy regime that corresponds to the 
transition between perturbative and non-perturbative QCD. Charmonium is one of the most simplest 
two-body systems in the field of hadronic physics. The charm quark 
is relatively heavy in mass, allowing theoretical analyses that are based on a 
non-relativistic framework with relativistic corrections such as spin-orbit and spin-spin forces.
Moreover, the charmonium states below the open-charm threshold are narrow as a consequence of the OZI suppression rule
and, hence, can easily be identified as sharp needles on top of a continuous background, thereby forming
ideal beacons of QCD. The level scheme of lower-lying bound charmonium states is very similar to that 
of positronium or the hydrogen atom. Figure~\ref{charmonium_spectrum}) gives an overview of the 
established mass spectrum of charmonium-like particles in the regime that is accessible by BESIII. 
The charmonium states below the open-charm threshold can be described fairly well in 
terms of heavy-quark potential models and rigorous lattice QCD calculations.
Precision measurements of the mass and width of the charmonium spectrum
give, therefore, access to the confinement potential in QCD.  
In addition, the various charmonium states with well-defined spin and parity
serve as ideal systems to study via their decay modes the validity of perturbative 
QCD and to probe the light-quark sector as well. For more details concerning
the underlying motivation of exploiting charmonium, I refer to Ref.~\cite{vol08}. 

\section{The BESIII experiment}

BEPCII is a two-ring e$^+$e$^-$ collider designed for a peak luminosity of 10$^{33}$~cm$^{-2}$s$^{-1}$ at a center-of-mass energy
of 3770~MeV. 
The cylindrical core of the BESIII detector consists of a helium-gas-based drift chamber, a plastic scintillator 
time-of-flight system, and a CsI(Tl) electromagnetic calorimeter, all enclosed in a superconducting solenoidal magnet providing 
a 1~T magnetic field. The solenoid is supported by an octagonal flux-return yoke with resistive plate counter muon identifier modules interleaved with steel. 
The charged particle and photon acceptance is 93\% of 4$\pi$, and the charged particle momentum and photon energy resolutions at 1~GeV are 0.5\% and 2.5\%, respectively. More details on the features and capabilities of BESIII are provided in Ref.~\cite{bes_bepc}. 
Both the BEPCII facility and the BESIII detector are major upgrades of the BESII detector and the BEPC accelerator. The first collisions with the complete 
setup took place in July of 2008. The first physics production runs started in the first half of 2009. Already during writing of this paper, the amount of data 
samples collected for the $J/\psi$, $\psi(3686)$, and $\psi(3770)$ is significantly larger than that obtained by the CLEO collaboration, 
thereby reaching a new world record in statistics. 

\section{Highlights in charmonium spectroscopy and decays}

In this paper, I discuss a few of the highlights that were obtained within the charmonium spectroscopy and decays program of BESIII. 
This program of BESIII resulted so far in a variety of papers that can be found in Refs.~\cite{hc_paper,gP_paper,PP_paper,matrix_paper,4pi_paper,gV_paper,chic_ppbarKK,chic_vectormesons,pseudoscalar_pipi,etac2s_vectormesons,psip_gchic,psip_getac,jpsi_raddecays,psip_jpsi_ngam,chic_gg,psip_getacp,chic_LLpipi,psi4010_etajpsi,psip_KKpi,chic_NNpi,hc_exclusive,psip_pijpsi,etac_gg,chic_etacpipi,chic_bb,psi_LLpi,etacp_KKpipipi}. 
The bulk of the activities below the open-charm threshold focuses on precision measurements of the properties of charmonium states, such as their masses, 
total widths, line shapes, and partial decay rates. This data provide an excellent basis to reveal the details of the confinement potential, to perform
perturbative QCD tests, and to study systematically non-perturbative, long-distance, contributions within a controlled environment.  
In addition, new states were discovered or confirmed with a large discovery potential above the open-charm threshold.

\subsection{The ground state and its first radial excitation of charmonium, $\eta_c(1S,2S)$}

Our present knowledge on the basic properties, such as mass and width, of the vector meson states ($J^{PC}=1^{--}$) of charmonium is excellent
due to the huge amount of data available. In contrary, the properties of the pseudo-scalar ($J^{PC}=0^{-+}$) charmonium states, 
including the ground state of charmonium and its radial excitation ($\eta_c(1S,2S)$), are poorly understood. For a large part this is due to 
the fact that these states cannot be populated directly via $e^+e^-$ annihilations. One can, however, study these states indirectly 
in $e^+e^-$ experiments via two-photon fusion, $B$ decays, or, as pursued with BESIII, via the magnetic-dipole (M1) 
transition $\psi(1S,2S)\rightarrow\gamma\eta_c$. 

\begin{figure} \includegraphics[width=\textwidth]{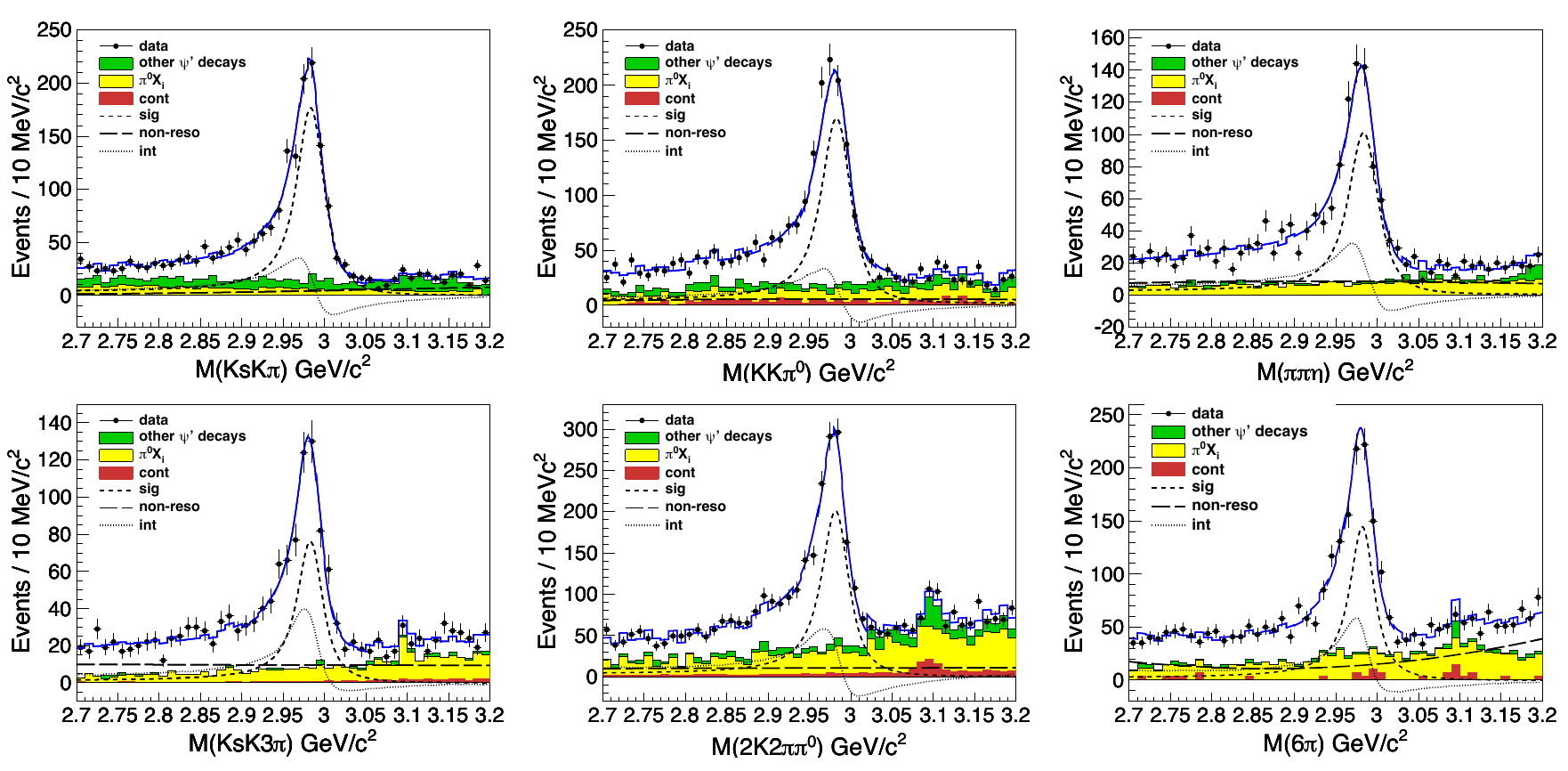}
\caption{The $M(X_i)$ invariant mass distributions for various decays, $X_i$, with the fit results superimposed.
Points are data and the various curves are the total fit results. Signals are shown as short-dashed lines;
the non-resonant components as long-dashed lines; and the interference between them as dotted lines. 
Shaded histograms are various backgrounds studied independently. The figure is taken from Ref.~\cite{psip_getac}.}
\label{fig_etac}
\end{figure}

Previously published measurements of the mass and width of the ground state, $\eta_c(1S)$, show large discrepancies 
among the various channels that were employed in the corresponding experiments~\cite{pdg2012}. The large statistics of BESIII allow to make a detailed 
study of the line shape of the $\eta_c(1S)$ via the channel $\psi(3686)\rightarrow\gamma\eta_c(1S)$ with the $\eta_c$ decaying in 
six exclusive channels. Figure~\ref{fig_etac} shows the result of the analysis. In all the decays, a clear signal from the
$\eta_c$ can be observed with a small amount of background. Note that the $\eta_c$ signal has an obviously asymmetric
shape: there is a long tail on the low-mass side while on the high-mass side the $\eta_c$ signal drops rapidly and
the data dips below the expected level of the smooth background, suggesting a possible interference with a non-resonant
contribution. A fit was performed for which the signal is described by a Breit-Wigner (BW) convolved with a resolution function
and including the possibility of an interference with a non-resonant background. The statistical significance of the interference
was found to be 15$\sigma$. Although, the nature of this interfering 
non-resonant contribution is not well understood yet, the high-statistics BESIII results show that a naive fit including 
a BW signal together with a smooth background assumption could result a misleading value for the mass and width of the
extracted $\eta_c$ resonance parameters. With the fitting procedure of BESIII a mass of $M$=2984.3$\pm$0.6(stat.)$\pm$0.6(syst.)~MeV/$c^2$
and a width of $\Gamma$=32.0$\pm$1.2(stat.)$\pm$1.0(syst.)~MeV were extracted. The BESIII results are consistent with
that from photon-photon fusion and $B$ decays. Using this measurement together with the world-average $J\/psi$ 
mass~\cite{pdg2012}, one obtains for the $S$-wave hyperfine mass splitting a value of 
$\Delta M_{hf}(1S)$=$M(J/\psi) - M(\eta_c)$=112.6$\pm$0.8~MeV/$c^2$, which
agrees well with recent lattice computations~\cite{lattice_burch,lattice_levkova,lattice_kawanai} and sheds light on the 
spin-dependent interactions in quarkonium states. More details can be found in Ref.~\cite{psip_getac}.     

\begin{figure} \includegraphics[width=\textwidth]{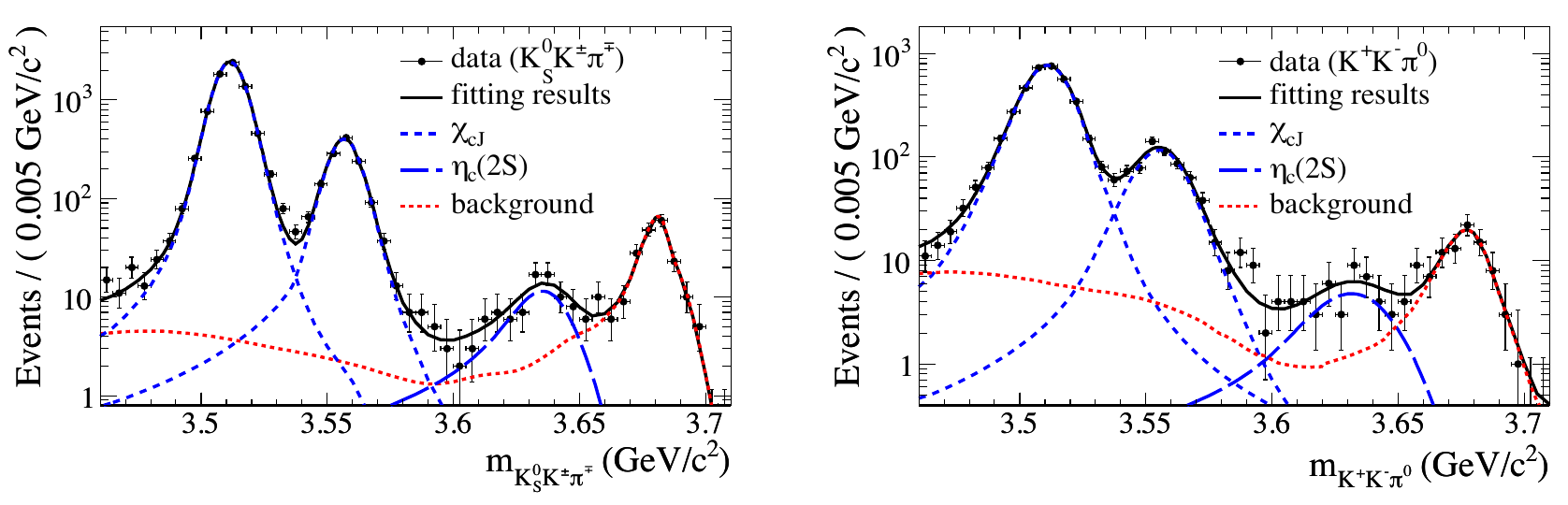}
\caption{The invariant mass spectrum for $K_S^0K^\pm\pi^mp$ (left panel), $K^+K^-\pi^0$ (right panel), and the
simultaneous likelihood fit to the three resonances and combined background sources. The figure is taken from Ref.~\cite{psip_getacp}.}
\label{fig_etacp}
\end{figure}

The experimental database on the first radially excited state of the ground state, $\eta_c(2S)$, is scarce. 
The $\eta_c(2S)$ was first observed by the Belle collaboration in the process $B^{\pm}\rightarrow K^\pm\eta_c(2S)$ with 
$\eta_c(2S)\rightarrow K_S^0K^\pm\pi^\mp$~\cite{belle_etacp} and later on confirmed in the two-photon production
of $K_S^0K^\pm\pi^mp$~\cite{babar_etacp,cleo_etacp} and in the double-charmonium production process
$e^+e^-\rightarrow J/\psi c\bar{c}$~\cite{babar_etacp2,belle_etacp2}. A controversial observation of the $\eta_c(2S)$
was made in the past with the crystal ball setup. This experiment performed an inclusive measurement of the energy 
spectrum of photons in $\psi(3686)$ decays and found a peak that corresponds to a missing mass of 3592$\pm$5~MeV/$c^2$ 
which they attributed to an observation of the $\eta_c(2S)$~\cite{cb_etacp}. Surprisingly, this measurement of the $\eta_c(2S)$
mass turns out to be significantly smaller than all the ones measured by Belle, BaBar, and CLEO. 
With the statistics collected with BESIII and the capability of the setup to study exclusive channels,
a first observation of the M1 transition was made possible in the reaction $\psi(3686)\rightarrow \gamma \eta_c(2S)$
with $\eta_c(2S)\rightarrow K\bar{K}\pi$. The results are depicted in Fig.~\ref{fig_etacp}. Besides the strong and well-understood
background signals from the $\chi_{c1,2}$ resonances and from initial and final-state radiative background processes, 
a clear peak can be observed in the expected mass region of the $\eta_c(2S)$. A simultaneous likelihood fit results
in a measurement of the $\eta_c(2S)$ mass of $M$=3637.6$\pm$2.9(stat.)$\pm$1.6(syst.)~MeV/$c^2$ and a width 
of $\Gamma$=16.9$\pm$6.4(stat.)$\pm$4.8(syst.)~MeV. In addition, a first measurement of the branching fraction of the
M1 transition $\psi(3686)\rightarrow\gamma\eta_c(2S)$ resulted in $B$=(6.8$\pm$1.1(stat.)$\pm$4.5(syst.))$\times$10$^{-4}$ where the relatively large systematic error stems from the uncertainty of the branching fraction of $\eta_c(2S)\rightarrow K\bar{K}\pi$ which was taken from a BaBar measurement~\cite{babar_kkpi}. More details of the BESIII observation
and corresponding measurements can be found in Ref.~\cite{psip_getacp}. More recently, the BESIII collaboration 
made an observation of the $\eta_c(2S)$ via the exclusive process 
$\psi(3686)\rightarrow \gamma K_S^0 K^+ \pi^- \pi^+ \pi^-$~\cite{etacp_KKpipipi}.
The extracted mass and width of this measurement were found to be within two and one standard deviations, respectively, 
from the measurements of the first $\eta_c(2S)$ observation of BESIII.  

\subsection{The singlet P-wave state of charmonium, $h_c$}

\begin{figure} \includegraphics[width=\textwidth]{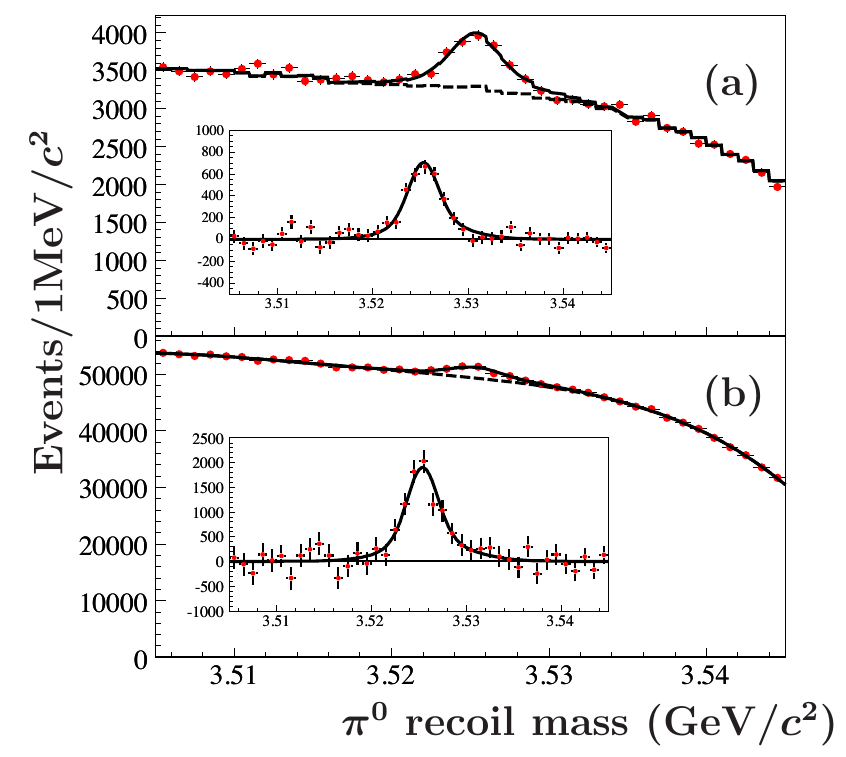}
\caption{(a) The $\pi^0$ recoil mass spectrum and fit for the E1-tagged analysis of 
$\psi(3686) \rightarrow \pi^0 h_c$, $h_c\rightarrow \gamma \eta_c$; (b) the $\pi^0$ recoil mass spectrum and fit for the 
inclusive analysis of $\psi(3686) \rightarrow \pi^0 h_c$. Fits are shown as solid lines, background as dashed lines. The insets show the 
background-subtracted spectra. Figures are taken from Ref.~\cite{hc_paper} and described there in more detail.}
\label{fig_hc}
\end{figure}

One of the important aspects related to quark confinement is the spin structure of the $q\bar q$ potential. 
The role of the spin-dependence in the hyperfine splitting of the $P$-waves is of particular interest. 
For this purpose, a precise measurement of the mass and decay channels of the singlet-$P$ resonance, $h_c$, is of extreme importance.
This state has been studied by the BESIII collaboration via the isospin-forbidden transition, $\psi(3686)\rightarrow \pi^0 h_c$. 
The results of this analysis are shown in Fig.~\ref{fig_hc}. Clear signals have been observed for this decay 
with and without the subsequent radiative decay, $h_c\rightarrow \gamma\eta_c$. This has led to a measurement of the mass and 
the total width of the $h_c$ of $M$=3525.40$\pm$0.13(stat.)$\pm$0.18(syst.)~MeV/$c^2$ and $\Gamma$=0.73$\pm$0.45(stat.)$\pm$0.28(syst.)~MeV ($<$1.44~MeV at 90\% C.L.), 
respectively. Furthermore, for the first time the branching fractions of the decays $\psi(3686)\rightarrow \pi^0 h_c$ 
and $h_c\rightarrow\gamma\eta_c$ were determined and found to be 
$\mathcal{B}(\psi(3686)\rightarrow \pi^0 h_c)=(8.4\pm1.3\pm1.0)\times 10^{-4}$ and $\mathcal{B}(h_c\rightarrow \gamma\eta_c)=(54.3\pm6.7\pm5.2)\%$, 
respectively. For a more detailed discussion, I refer to Ref.~\cite{hc_paper}.

\begin{figure} \includegraphics[width=\textwidth]{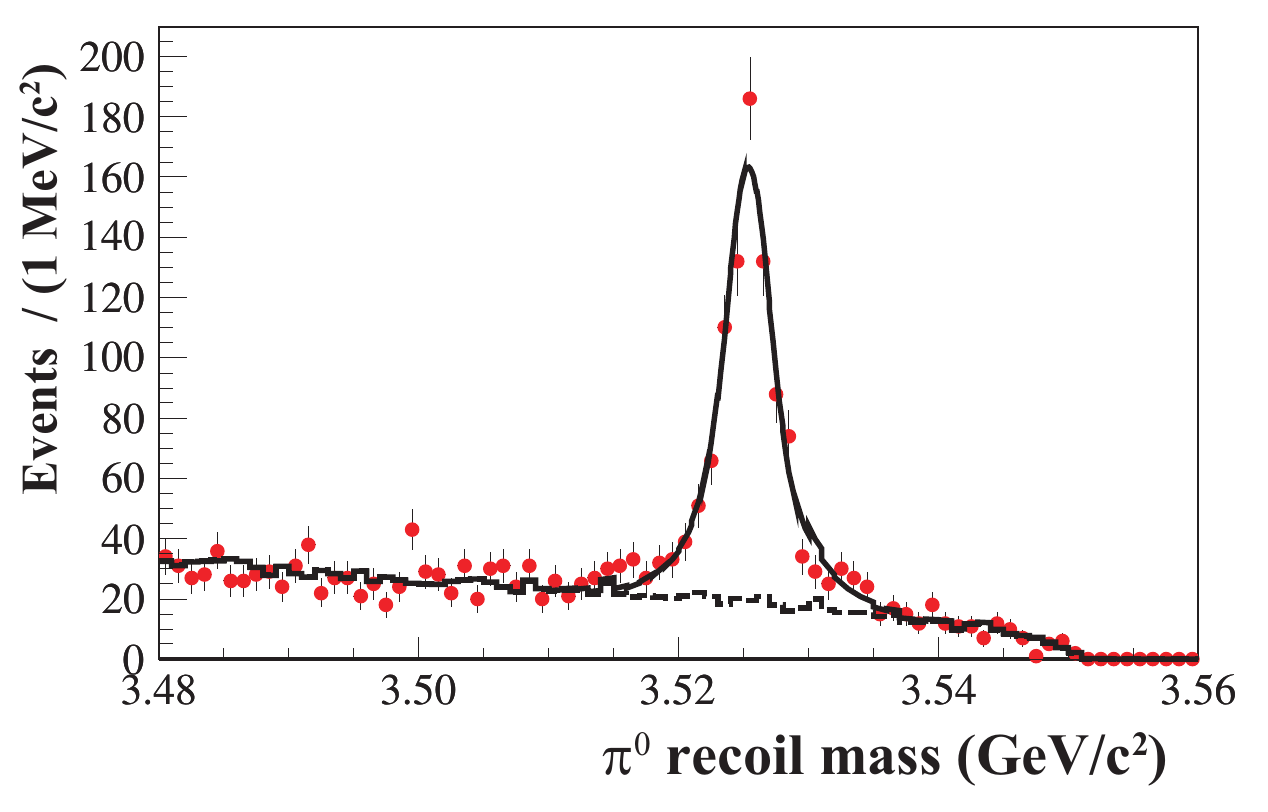}
\caption{The $\pi^0$ recoil mass spectrum in $\psi(3686)\rightarrow \pi^0 h_c$, $h_c\rightarrow \gamma \eta_c$, 
$\eta_c\rightarrow X_i$ summed over 16 various final states $X_i$. The dots with error bars represent data, 
the solid line shows the total fit function, and the dashed line is the background component of the fit.
The figure is taken from Ref.~\cite{hc_exclusive}.}
\label{fig_hc2}
\end{figure}

More recently, the process $\psi(3686)\rightarrow\pi^0 h_c$ followed by $h_c\rightarrow \gamma\eta_c$ was
re-analyzed by BESIII with an exclusive-reconstruction technique. For this, the signal-to-background ratio was improved
drastically by selecting 16 exclusive hadronic decays of the $\eta_c$. The results of this work are shown
in Fig.~\ref{fig_hc2}. A clear signal from the $h_c$ can be observed with an improved significance with respect
to the data shown in Fig.~\ref{fig_hc}. This analysis resulted in the world's best determination of the $h_c$ mass of
$M$=3525.31$\pm$0.11(stat.)$\pm$0.14(syst.)~MeV/$c^2$ and a width of $\Gamma$=0.70$\pm$0.28(stat.)$\pm$0.22(syst.)~MeV.
With this mass measurement together with the very-well determined masses of the triplet $P$-waves in charmonium, $\chi_{c0,1,2}$, 
one finds for the $1P$-wave hyperfine mass splitting a value of 
$\Delta M_{hf}$$\equiv$$<M(1^3P)>-M(1^1P_1)$ = $-0.01$$\pm$0.11(stat.)$\pm$0.15(syst.)~MeV/$c^2$,
which is consistent with the absence of a strong spin-spin interaction. Details of the analysis and results can be found in Ref.~\cite{hc_exclusive}.

\subsection{Two- and three-photon decays of charmonium}

Decays of positronium to more than one photon are regarded as ideal test-beds for quantum electrodynamics (QED). 
Similarly, the multi-photon decays of charmonium states can serve as promising processes to test the
strong-force equivalent of QED, namely QCD~\cite{vol08}. In general, multi-photon decays provide a clean tool
to study the nature of interquark forces via the annihilation of the $c$ and $\bar{c}$ quarks. 
For this reason, the BESIII collaboration has studied various multi-photon decays of charmonium states below 
the open-charm threshold. 

A precision measurement of the two-photon widths of the triplet $P$-waves in charmonium via the processes 
$\chi_{c0,2}\rightarrow\gamma\gamma$ have been performed in analog to the corresponding triplet states
of positronium. In lowest order, for both positronium and charmonium, the ratio of the two-photon decays
$R_{th}^{(0)}=\frac{\Gamma(^3P_2\rightarrow \gamma\gamma)}{\Gamma(^3P_0\rightarrow\gamma\gamma)}=4/15\approx 0.27$~\cite{vol08}.
Any discrepancy from this simple lowest order prediction can arise due to QCD radiative corrections
and relativistic corrections. Hence, a measurement on $R$ and a comparison with 
theory, allows to systematically test such corrections and will, thereby, guide
the development of QCD theory. 

\begin{figure} \includegraphics[width=\textwidth]{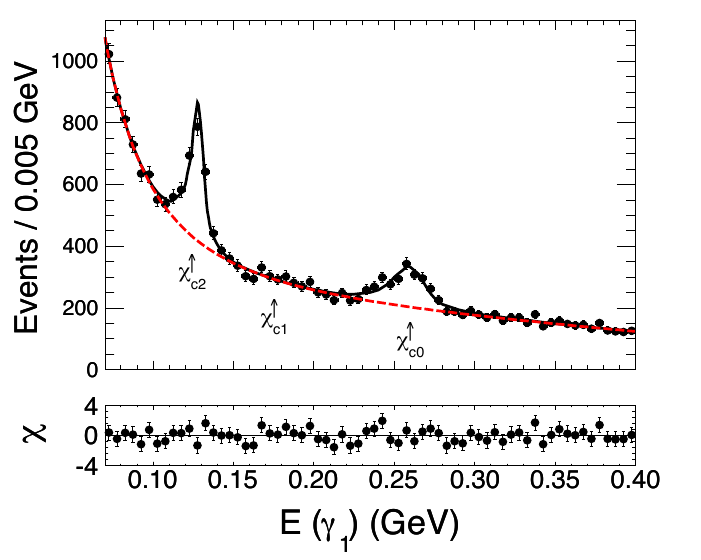}
\vspace*{-1cm}
\caption{The upper plot shows the fitted $E_{\gamma_1}$ spectrum for the $\psi(3686)$
data sample. The expected positons of $E_{\gamma_1}$ from $\chi_{c0,1,2}$ are
indicated by arrows. The dashed curve shows the background line shape and the
solid curve the result of the full fit. The lower plot illustrates the
quality of the fit by plotting the number of standard deviations, $\chi$, 
of data points with respect to the fitted curve. 
The figure is taken from Ref.~\cite{chic_gg} and described there in more detail.}
\label{fig_chic_2g}
\end{figure}

\begin{sloppypar}
With BESIII, one can access the $\chi_c$ states via the electric-dipole (E1)
transition $\psi(3686)\rightarrow \gamma\chi_c$, and look for its subsequent decay into
photons, e.g. $\chi_c\rightarrow \gamma\gamma$. Figure~\ref{fig_chic_2g} shows the results 
obtained with BESIII. The figure depicts the energy distribution of the candidate E1-transition
photon, $\gamma_1$, whereby additionally two photons have been registered. Clearly, one observes strong signals
from the processes $\chi_{c0,2}\rightarrow\gamma\gamma$. The decay $\chi_{c1}\rightarrow\gamma\gamma$
cannot be observed, which can be explained by the Landau-Yang theorem. BESIII finds branching fractions
of $\mathcal{B}(\chi_{c0}\rightarrow \gamma\gamma)$=(2.24$\pm$0.19(stat.)$\pm$0.15(syst.))$\times$10$^{-4}$ and
$\mathcal{B}(\chi_{c2}\rightarrow \gamma\gamma)$=(3.21$\pm$0.18(stat.)$\pm$0.22(syst.))$\times$10$^{-4}$, which
both agrees with the less-precise results from the CLEO experiment~\cite{cleo_chicgg}.
With these branching fractions, one obtains a value of $R=0.27\pm 0.04$, where various systematic
errors cancel out. Although, the experimental measurement of $R$ agrees with the lowest-order prediction
of $R_{th}^{(0)}$, it deviates significantly with the first-order predictions including radiative corrections,
$R_{th}^{(1)}=0.116\pm 0.010$~\cite{vol08}. In addition, the BESIII data allow for the first time a helicity
amplitude analysis for the decay $\psi(3686)\rightarrow \gamma\chi_{c2}$ with $\chi_{c2}\rightarrow\gamma\gamma$.
The ratio of the two-photon partial widths for the helicity-zero and  helicity-two components in the decay
$\chi_{c2}\rightarrow \gamma\gamma$ is determined to be $f_{0/2}$=0.00$\pm$0.02(stat.)$\pm$0.02(syst.).
Note that the helicity-zero component is highly suppressed, which observation is consistent with relativistic 
potential calculations. For a detailed description, I refer to~\cite{chic_gg}.
\end{sloppypar}

\begin{figure} \includegraphics[width=\textwidth]{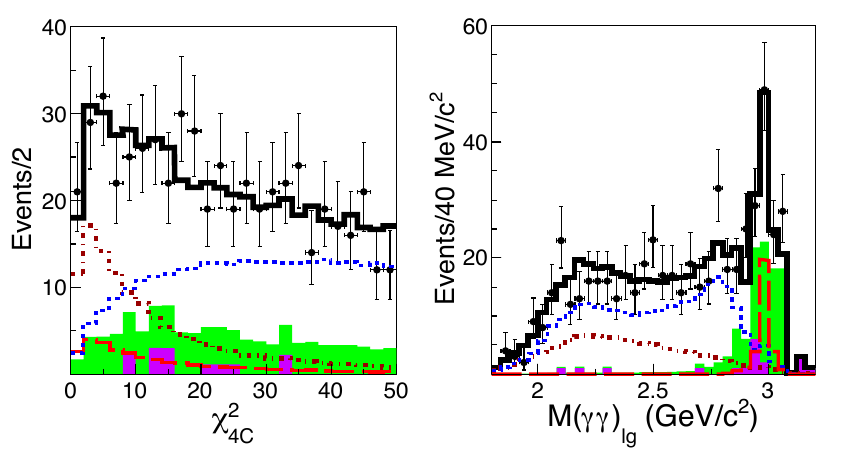}
\vspace*{-1cm}
\caption{The left panel shows the chisquare distribution of a four-constraint fit, $\chi_{4C}^2$
and the right panel shows the two-photon invariant mass, $M(\gamma\gamma)_{\rm lg}$, both for the
process $\psi(3686)\rightarrow\pi^+\pi^- J/\psi$ with $J/\psi\rightarrow 3\gamma$. The data are indicated
as points with error bars and the results of a multi-dimensional fit are shown as thick solid lines.
The (dark red) dotted-dashed, (red) dashed and (blue) dotted lines
show contributions from $J/\psi\rightarrow 3\gamma$, $J/\psi\rightarrow \gamma\eta_c$, and $J/\psi\rightarrow \gamma\pi^0\pi^0$,
respectively. The stacked histogram represents the backgrounds from $J/\psi\rightarrow \gamma\pi^0/\eta/\eta^\prime$
(light shaded and green) and non-$J/\psi$ decays (dark shaded and violet). This figure is taken from Ref.~\cite{etac_gg} and
described there in more detail.}
\label{fig_etac_2g}
\end{figure}

The two-photon decay of the $\eta_c(1S)\rightarrow\gamma\gamma$ is experimentally very challenging
and it is one of the key benchmarks for BESIII to demonstrate its capabilities. 
Since the $\eta_c(1S)$ cannot be created directly in
$e^+e^-$ annihilations, one has to exploit the suppressed radiative M1 transitions of the vector-meson states
of charmonium. With BESIII, the $\eta_c(1S)\rightarrow \gamma\gamma$ channel has been studied via the 
cascade process $\psi(3686)\rightarrow \pi^+\pi^- J/\psi$ with $J/\psi\rightarrow\gamma\eta_c(1S)$ and, eventually, 
$\eta_c(1S)\rightarrow \gamma\gamma$. Although, BESIII also has taken a large data sample at the $J/\psi$ mass, 
the additional $\psi(3686)\rightarrow \pi^+\pi^- J/\psi$ has the advantage of tagging exclusively on the $J/\psi$.
Also note that the branching fraction of the $\pi^+\pi^-$ is relatively large (34\%). Besides the two-photon
decay of the $\eta_c(1S)$, the same study also provides access to the direct three-photon transition of the 
$J/\psi\rightarrow 3\gamma$. Note that the charge conjugation of the $J/\psi$ does not allow 
an annihilation of the $c\bar{c}$ into two photons. The photon energy spectrum of the $J/\psi\rightarrow 3\gamma$
can as well be used to study the internal structure of the $J/\psi$, since the value of the photon energy 
is a measure for the distance between the quarks. 

Figure~\ref{fig_etac_2g} shows a key ploy of the $J/\psi\rightarrow 3\gamma$ analysis. The figure shows the
projections of a two-dimensional maximum likelihood fit that was performed on the distributions with respect to
the two-photon invariant mass, $M(\gamma\gamma)_{\rm lg}$, of the two photons with the highest energy, 
and the chisquare of the four-constraint kinematic fit, $\chi_{4C}^2$. The fit was used to estimate the
yields of the direct $J/\psi\rightarrow 3\gamma$ process and the cascade channel
$J/\psi\rightarrow \gamma (\gamma\gamma)_{\eta_c(1S)}$. Although the data suffer from well-understood 
background sources from the decays $J/\psi\rightarrow \gamma\pi^0\pi^0$ and $J/\psi\rightarrow \gamma\pi^0/\eta/\eta^\prime$, 
the fit provides a clear evidence of the presence of contributions from the two- and three-photon decays of the
$\eta_c(1S)$ and $J/\psi$, respectively. The branching fraction of the direct decay of $J/\psi\rightarrow 3\gamma$ is measured
to be $\mathcal{B}(J/\psi\rightarrow 3\gamma)$=(11.3$\pm$1.8$\pm$2.0)$\times$10$^{-6}$, which is consistent with and more accurate than
the result from CLEO~\cite{cleo_jpsi_3g}. The measured relative branching fraction 
$\mathcal{B}(J/\psi\rightarrow 3\gamma)/\mathcal{B}(J/\psi\rightarrow e^+e^-)$ is incompatible with expectations including first-order
QCD corrections, pointing to a need for further improvements of the QCD radiative and relativistic corrections. 
More details can be found at Ref.~\cite{etac_gg}.

\subsection{Charmed-meson loops and isospin-violating transitions}

Isospin is known to be a good symmetry in the charmonium states below the
$D\bar{D}$-production threshold. The decay rates of isospin-symmetry breaking modes are
in general found to be very small. For example, the branching fraction of the
isospin-violating transition $\psi(3686)\rightarrow \pi^{0} J/\psi$ is known to much smaller 
than the branching fraction to other hadronic transitions such as the 
$\psi(3686)\rightarrow \pi^{+}\pi^{-} J/\psi$ process.

Although the isospin-breaking is found to be very small for the light charmonium states,
the mysterious $X(3872)$ resonance above the $D\bar{D}$ threshold decays predominantly
via the transition $X(3872)\rightarrow \pi^{+}\pi^{-} J/\psi $ where the
invariant-mass spectrum of the $\pi^{+}\pi^{-}$ pair shows a clear $\rho$ signature~\cite{X3872Belle2003isospin}
and, hence, is compatible with an isospin-violating decay.
A possible scenario is that the $X(3872)$ is a molecular state composed of a bound $D^{*0}$-$\bar{D}^0$
meson pair. Such an explanation is particularly popular since the mass of the
$X(3872)$ is only slightly less than the sum of the $D^0$ and $D^{*0}$ masses, pointing to
a state that is weakly bound by the exchange of a color-neutral meson, similar to the deuteron. 
Moreover, in such a scenario, the strong isospin-breaking decay rate of the $X(3872)$ could be explained 
by the large mass gap between the $D^{0*}-\bar{D}^0$ and the $D^{*+}-D^-$ ($D^+-D^{*-}$) thresholds. 
A better understanding of the isospin-breaking mechanism in a controlled system,
such as charmonium, could be crucial to shed light on the true nature of the $X(3872)$.

A non-relativistic effective-field theoretical (NREFT) study by the J\"{u}lich and IHEP groups showed that 
intermediate (virtual) charmed-meson loops can be a dominant source for the isospin breaking. 
Detailed studies of different isospin-violating transitions in charmonium below the open-charm threshold and the effect of 
virtual charmed-meson loops on the widths of the transitions are described in \cite{Hanhart2009,Hanhart2010}.  
These NREFT calculations are based on a first estimate exploiting diagrams involving the lowest-lying pseudoscalar 
and vector charmed mesons following heavy-quark symmetry and chiral symmetry. With a complete effective field theory 
including Goldstone bosons, charmonia, and charmed mesons as the
degrees of freedom, it would be possible in the future to make a rigorous interpretation of the nature
of the $XYZ$ states, such as the $X(3872)$, and, moreover, to extract the light-quark masses from quarkonia decays. 
Currently, for such a theory, quantitative predictions on individual branching fractions of isospin-forbidden decays 
of charmonium are difficult to make, because the information on the coupling constants $f_{\psi DD}$ between different charmonium 
states and $DD$-mesons is limited. The theory requires constraints from experimental data, in particular from 
measurements of decay rates of various isospin-violating transitions in charmonium. 

\begin{figure} \includegraphics[width=\textwidth]{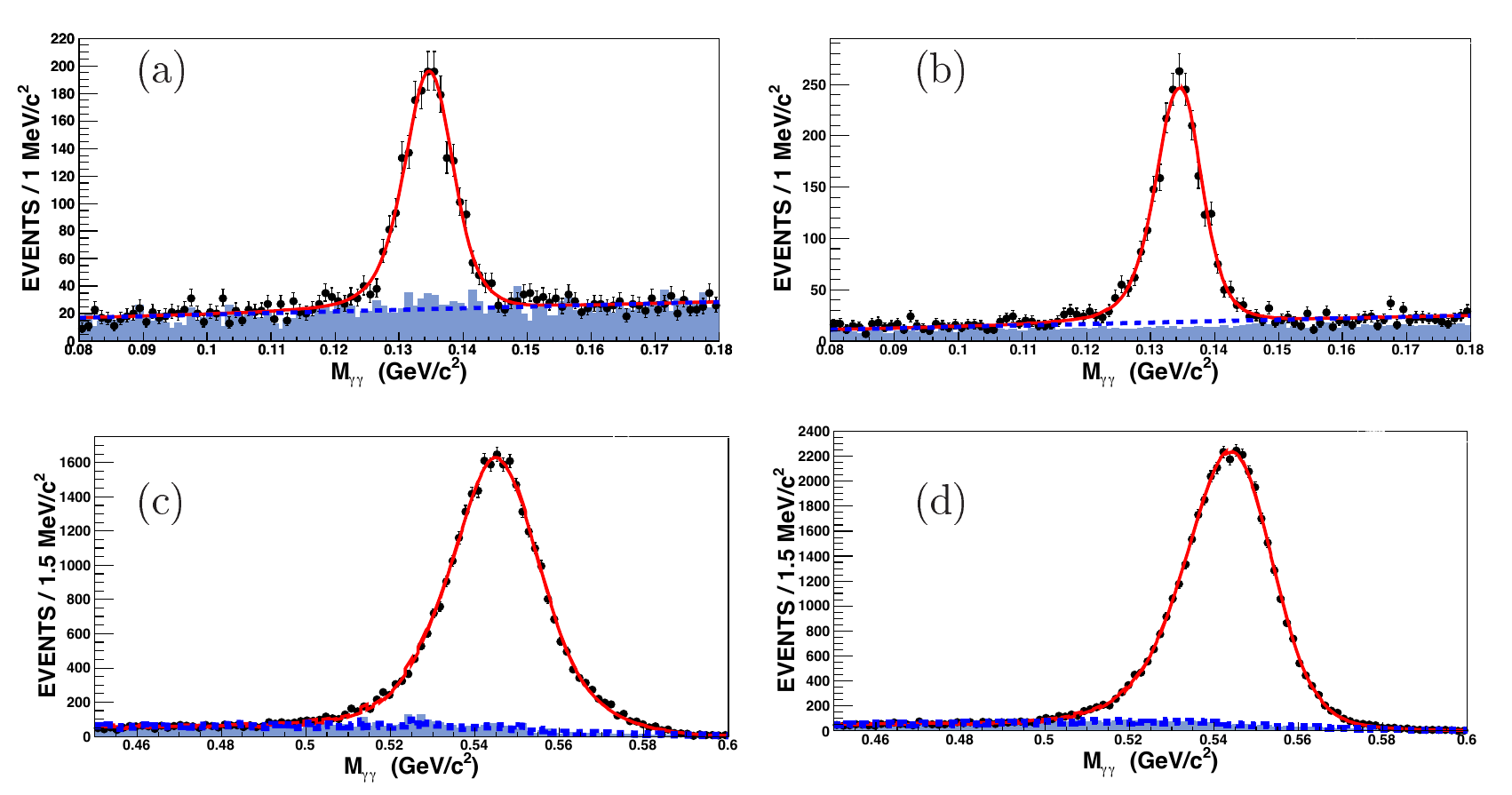}
\vspace*{-0.8cm}
\caption{Two-photon invariant-mass ($M_{\gamma\gamma}$) distributions and fit results. (a) $\psi(3686)\rightarrow \pi^0 J/\psi$
with $J/\psi\rightarrow e^+e^-$, (b) $\psi(3686)\rightarrow\pi^0 J/\psi$ with $J/\psi\rightarrow \mu^+\mu^-$,
(c) $\psi(3686)\rightarrow \eta J/\psi$ with $J/\psi\rightarrow e^+e^-$, (d) $\psi(3686)\rightarrow\eta J/\psi$ 
with $J/\psi\rightarrow \mu^+\mu^-$, where the points with error bars are data, and the solid (red) curves are the results
of the total fit, and the dashed curves are the fitted background shapes. The hatched histograms represent background events
obtained from Monte Carlo simulations and from analyses of sidebands in the $J/\psi$-mass spectrum. 
This figure is taken from Ref.~\cite{psip_pijpsi} and described there in more detail.}
\label{fig_psip_pijpsi}
\end{figure}

The BESIII collaboration has started a campaign to provide the best measurements of the isospin-forbidden transition
rates below the open-charm threshold. Besides the decay $\psi(3686)\rightarrow\pi^0 h_c$, presented earlier in this paper,
the most precise measurement of the branching fraction of the isospin-forbidden decay $\psi(3868)\rightarrow \pi^0 J/\psi$ with 
$J/\psi\rightarrow l^+l^-$ was recently published by the BESIII collaboration~\cite{psip_pijpsi}. In addition, 
the branching fraction of the isospin-allowed decay $\psi(3868)\rightarrow \eta J/\psi$ was extracted. 
Figure~\ref{fig_psip_pijpsi} depicts the two-photon invariant mass spectra that were analyzed to extract the corresponding
branching fractions. In all channels, a clear signal can be observed on top of a smooth background. 
Using this data, BESIII obtained the branching fractions $\mathcal{B}(\psi(3686)\rightarrow \pi^0 J/\psi)$=(1.26$\pm$0.02(stat.)$\pm$
0.03(syst.))$\times$10$^{-3}$ and $\mathcal{B}(\psi(3686)\rightarrow \eta J/\psi)$=(33.75$\pm$0.17(stat.)$\pm$0.86(syst.))$\times$10$^{-3}$.
The branching fraction ratio
$R=\mathcal{B}(\psi(3686)\rightarrow \pi^0 J/\psi)/\mathcal{B}(\psi(3686)\rightarrow \eta J/\psi)$ was determined to be 
(3.74$\pm$0.06(stat.)$\pm$0.04(syst.))$\times$10$^{-2}$. These results indicate that contributions from 
charmed-meson loops are a possible mechanism for the dominant source of isospin violation~\cite{Hanhart2009}. 

\subsection{Terra incognita}

The activities below the open-charm threshold are mainly focussed towards precision charmonium studies
and light-hadron spectroscopy. The physics case above the open-charm threshold is for a large part devoted to new discoveries. 
One of the current interests of BESIII is to study the nature of the so-called $Y(4260)$. This state was discovered 
via the initial-state-radiation (ISR) process $e^+e^-\rightarrow \gamma_{\rm ISR}\pi^+\pi^- J/\psi$ 
by the BaBar collaboration~\cite{babar_y4260,babar_y42602} and confirmed by 
CLEO~\cite{cleo_y4260,cleo_y42602}, and Belle~\cite{belle_y4260}. Unlike other 
charmonium states with the same quantum numbers in the same mass region,
such as the $\psi(4040)$,$\psi(4160)$, and $\psi(4415)$, the $Y(4260)$ state does not have a natural place within
the quark model of charmonium. Furthermore, while being well above the $D\bar{D}$ threshold, the $Y(4260)$ shows
strong coupling to the $\pi^+\pi^- J/\psi$ final state, but relatively small coupling to open charm decay modes.
These aspects could point to an unconventional state, for instance, a charmonium hybrid.

\begin{figure} \includegraphics[width=\textwidth]{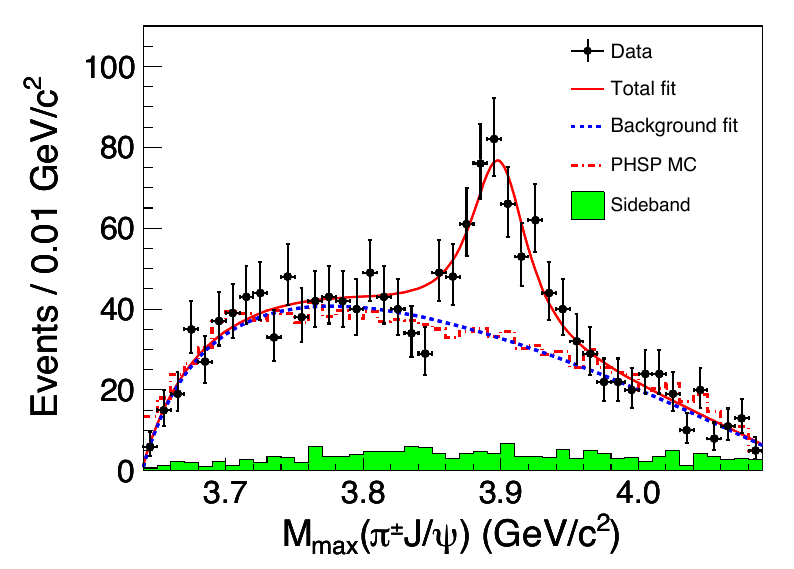}
\vspace*{-0.8cm}
\caption{The invariant-mass distribution of the $\pi^\pm J/\psi$ pair in $e^+e^-\rightarrow \pi^+\pi^- J/\psi$
at a center-of-mass energy of 4.26~GeV. For each event, only the maximum mass is taken out of the two combination
from the two charged pions. Dots with error bars are data; the red solid curve shows the total fit; the red 
dot-dashed histogram shows the result of a phase space Monte Carlo simulation; and the green shaded histogram 
shows the normalized $J/\psi$ sideband events. This figure is taken from Ref.~\cite{bes3_x3900} and described 
there in more detail.}
\label{fig_zc4260}
\end{figure}

Starting from the end of 2012, data were taken at a center-of-mass energy of $\sqrt{s}$=4260~MeV. The first 525~pb$^{-1}$
of luminosity at this energy gave already unexpected results. To the surprise of the BESIII collaboration, a new state
was discovered in the invariant-mass spectrum of the $\pi^\pm J/\psi$ pair~\cite{bes3_x3900} (see Fig.~\ref{fig_zc4260}). 
The structure has a mass of (3899$\pm$3.6(stat.)$\pm$4.9(syst.))~MeV/$c^2$ and a width of (46$\pm$10(stat.)$\pm$20(syst.))~MeV.
The intriguing aspect of the state is that it couples to charmonium and that it has an electric charge, which is suggestive 
of a state containing more quarks than just a charm and an anti-charm quark. Also the Belle collaboration and the CLEO experiment
reported short after the announcement by BESIII on a new observation of a charged charmonium-like
state at a similar mass~\cite{belle_x3900,cleo_x3900}. Likely, many new discoveries and insights are to be expected in the 
upcoming years.

\section{Summary}

Since the discovery of charmonium in November 1974, the field of hadron and particle physics exploiting systems made from
charm quarks has made a tremendous progress. Meanwhile, all the expected charmonium states below the open-charm threshold
have been discovered, and for most of them, the basic resonance parameters are well studied with high statistics data from
experiments such as Belle, BaBar, CLEO, and BESIII. Also, from the theoretical side, progress has been made. 
Precise calculations based on first principles are fastly becoming available, such as lattice QCD and effective-field 
theoretical approaches. The future challenges in charmonium spectroscopy are in finding and understanding the states 
above the open-charm threshold, in particular the recently discovered $XYZ$ states and the many so-far unobserved
higher lying charmonium states.

This paper reviews a few of the highlights obtained with BESIII. In particular, the paper summarizes the
measurements that were performed on the pseudo-scalar charmonium states, the singlet $P$-wave, two- and three photon
decays, isospin-violating transitions, and the recent BESIII activities at a center-of-mass energy of $\sqrt{s}$=4.26~GeV.
BESIII is presently one of the leading experiments in the field of charmonium spectroscopy and decays and has collected
a world-record on data at various energies, resulting in precision studies and new discoveries.  

\section*{Acknowledgments}

I thank the organizers of the Bormio winter meeting for giving me the opportunity to present of few of the exciting 
new results of BESIII. Furthermore, I thank all the members of the BESIII collaboration for their excellent achievements 
that has contributed to this paper. This work is supported by the University of Groningen (RuG) and the 
Helmholtzzentrum f\"ur Schwerionenforschung GmbH (GSI).


\begin{thebibliography}{99}

\bibitem{vol08} M.~B.~Voloshin, \emph{Charmonium}, Prog. Part. Nucl. Phys. {\bf 61}, 455 (2008) [{\tt hep-ph/0711.4556}].
\bibitem{bes_bepc} M.~Ablikim {\it et al.} (BESIII Collaboration), \emph{Design and construction of the BESIII detector}, Nucl. Instrum. Meth. A {\bf 614}, 345 (2010) [{\tt ins-det/0911.4960}].
\bibitem{hc_paper} M.~Ablikim {\it et al.} (BESIII Collaboration), \emph{Measurements of $h_c(^1P_1)$ in $\psi^\prime$ Decays}, Phys. Rev. Lett. {\bf 104}, 132002 (2010) [{\tt hep-ex/1002.0501}].
\bibitem{gP_paper} M.~Ablikim {\it et al.} (BESIII Collaboration), \emph{Evidence for $\psi^\prime$ decays into $\gamma\pi^0$ and $\gamma\eta$}, Phys. Rev. Lett. {\bf 105}, 261801 (2010) [{\tt hep-ex/1011.0885}].
\bibitem{PP_paper} M.~Ablikim {\it et al.} (BESIII Collaboration), \emph{Branching fraction measurements of $\chi_{c0}$ and $\chi_{c2}$ to $\pi^0\pi^0$ and $\eta\eta$},  Phys. Rev. D {\bf 81}, 052005 (2010) [{\tt hep-ex/1001.5360}].
\bibitem{matrix_paper} M.~Ablikim {\it et al.} (BESIII Collaboration), \emph{Measurement of the matrix element for the decay $\eta^{\prime} \to \eta \pi^+\pi^-$}, Phys. Rev. D {\bf 83}, 012003 (2011) [{\tt hep-ex/1012.1117}].
\bibitem{4pi_paper} M.~Ablikim {\it et al.} (BESIII Collaboration), \emph{First observation of the decays $\chi_{cJ} \to \pi^0 \pi^0 \pi^0 \pi^0$}, Phys. Rev. D {\bf 83}, 012006 (2011) [{\tt hep-ex/1011.6556}].
\bibitem{gV_paper} M.~Ablikim {\it et al.} (BESIII Collaboration), \emph{Study of $\chi_{cJ}$ radiative decays into a vector meson}, Phys. Rev. D {\bf 83}, 112005 (2011) [{\tt hep-ex/1103.5564}].
\bibitem{chic_ppbarKK} M.~Ablikim {\it et al.} (BESIII Collaboration), \emph{Observation of $\chi_{cJ}$ decaying into the $p\bar{p}K^+K^-$ final state}, Phys. Rev. D {\bf 83}, 112009 (2011) [{\tt hep-ex/1103.2661}].
\bibitem{chic_vectormesons} M.~Ablikim {\it et al.} (BESIII Collaboration), \emph{Observation of $\chi_{c1}$ decays into vector meson pairs $\phi\phi$, $\omega\omega$ and $\omega\phi$}, Phys. Rev. Lett. {\bf 107}, 092001 (2011) [{\tt hep-ex/1104.5068}].
\bibitem{pseudoscalar_pipi} M.~Ablikim {\it et al.} (BESIII Collaboration), \emph{Search for CP and P violating pseudoscalar decays into $\pi\pi$}, Phys. Rev. D {\bf 84}, 032006 (2011) [{\tt hep-ex/1106.5118}].
\bibitem{etac2s_vectormesons} M.~Ablikim {\it et al.} (BESIII Collaboration), \emph{Search for $\eta_c(2S)$ decays into vector meson pairs}, Phys. Rev. D {\bf 84}, 091102(R) (2011) [{\tt hep-ex/1110.0949}].
\bibitem{psip_gchic} M.~Ablikim {\it et al.} (BESIII Collaboration), \emph{Higher-order multipole amplitude measurement in $\psi^\prime$ decays into $\gamma\chi_{cJ}$}, Phys. Rev. D {\bf 84}, 092006 (2011) [{\tt hep-ex/1110.1742}].
\bibitem{psip_getac} M.~Ablikim {\it et al.} (BESIII Collaboration), \emph{Measurements of the mass and width of the $\eta_c$ using $\psi^\prime\rightarrow\gamma\eta_c$}, Phys. Rev. Lett. {\bf 108}, 222002 (2012) [{\tt hep-ex/1111.0398}].
\bibitem{jpsi_raddecays} M.~Ablikim {\it et al.} (BESIII Collaboration), \emph{Search for a light Higgs-like boson $A^0$ in $J/\psi$ radiative decays}, Phys. Rev. D {\bf 85}, 092012 (2012) [{\tt hep-ex/1111.2112}].
\bibitem{psip_jpsi_ngam} M.~Ablikim {\it et al.} (BESIII Collaboration), \emph{Evidence for the direct two-photon transition from $\psi^\prime$ to $J/\psi$}, Phys. Rev. Lett. {\bf 109}, 172002 (2012) [{\tt hep-ex/1204.0246}].
\bibitem{chic_gg} M.~Ablikim {\it et al.} (BESIII Collaboration), \emph{Two-photon widths of the $\chi_{c0,2}$ states and helicity analysis for $\chi_{c2}$ to $\gamma\gamma$}, Phys. Rev. D {\bf 85}, 112008 (2012) [{\tt hep-ex/1205.4284}].
\bibitem{psip_getacp} M.~Ablikim {\it et al.} (BESIII Collaboration), \emph{First observation of the M1 transition $\psi^\prime\rightarrow\gamma\eta_c^\prime$}, Phys. Rev. Lett. {\bf 109}, 042003 (2012) [{\tt hep-ex/1205.5103}].
\bibitem{chic_LLpipi} M.~Ablikim {\it et al.} (BESIII Collaboration), \emph{Observation of $\chi_{cJ}$ decays to $\Lambda\bar{\Lambda}\pi^+\pi^-$}, Phys. Rev. D {\bf 86}, 052004 (2012) [{\tt hep-ex/1207.5646}].
\bibitem{psi4010_etajpsi} M.~Ablikim {\it et al.} (BESIII Collaboration), \emph{Observation of $e^+e^-\rightarrow \eta J/\psi$ at center-of-mass energy $\sqrt{s}$=4.009~GeV}, Phys. Rev. D {\bf 86}, 071101(R) (2012) [{\tt hep-ex/1208.1857}].
\bibitem{psip_KKpi} M.~Ablikim {\it et al.} (BESIII Collaboration), \emph{Experimental study of $\psi^\prime$ decays to $K^+K^-\pi^0$ and $K^+K^-\eta$}, Phys. Rev. D {\bf 86}, 072011 (2012) [{\tt hep-ex/1208.2320}].
\bibitem{chic_NNpi} M.~Ablikim {\it et al.} (BESIII Collaboration), \emph{Measurement of $\chi_{cJ}$ decaying into $pn\pi^-$ and $pn\pi^-\pi^0$}, Phys. Rev. D {\bf 86}, 052011 (2012) [{\tt hep-ex/1208.3721}].
\bibitem{hc_exclusive} M.~Ablikim {\it et al.} (BESIII Collaboration), \emph{Study of $\psi(3686)\rightarrow\pi^0 h_c$, $h_c\rightarrow \gamma\eta_c$ via $\eta_c$ exclusive decays}, Phys. Rev. D {\bf 86}, 092009 (2012) [{\tt hep-ex/1209.4963}].
\bibitem{psip_pijpsi} M.~Ablikim {\it et al.} (BESIII Collaboration), \emph{Precision measurements of branching fractions for $\psi(3686)\rightarrow\pi^0 J/\psi$ and $\eta J/\psi$}, Phys. Rev. D {\bf 86}, 092008 (2012) [{\tt hep-ex/1210.3746}].
\bibitem{etac_gg} M.~Ablikim {\it et al.} (BESIII Collaboration), \emph{Evidence for $\eta_c\rightarrow \gamma\gamma$ and measurement of $J/\psi\rightarrow 3\gamma$}, Phys. Rev. D {\bf 87}, 032003 (2013) [{\tt hep-ex/1208.1461}].
\bibitem{chic_etacpipi} M.~Ablikim {\it et al.} (BESIII Collaboration), \emph{Search for hadronic transition $\chi_{cJ}\rightarrow \eta_c\pi^+\pi^-$ and observation of $\chi_{cJ}\rightarrow K\bar{K}\pi\pi\pi$}, Phys. Rev. D {\bf 87}, 012002 (2013) [{\tt hep-ex/1208.4805}].
\bibitem{chic_bb} M.~Ablikim {\it et al.} (BESIII Collaboration), \emph{Measurements of baryon pair decays of $\chi_{cJ}$ mesons}, Phys. Rev. D {\bf 87}, 032007 (2013) [{\tt hep-ex/1211.2283}].
\bibitem{psi_LLpi} M.~Ablikim {\it et al.} (BESIII Collaboration), \emph{Measurements of the branching fractions for $J/\psi$ and $\psi(3686)\rightarrow \Lambda\bar{\Lambda}\pi^0$ and $\Lambda\bar{\Lambda}\eta$}, Phys. Rev. D {\bf 87}, 052007 (2013) [{\tt hep-ex/1211.4682}].
\bibitem{etacp_KKpipipi} M.~Ablikim {\it et al.} (BESIII Collaboration), \emph{Evidence for $\eta_c(2S)$ in $\psi(3686)\rightarrow \gamma K_S K^+ \pi^-\pi^+\pi^-$}, Phys. Rev. D {\bf 87}, 052005 (2013) [{\tt hep-ex/1301.1476}].
\bibitem{pdg2012} J.~Beringer {\it et al.} (Particle Data Group), \emph{Review of Particle Physics}, Phys. Rev. D {\bf 86}, 010001 (2012).
\bibitem{lattice_burch} T.~Burch {\it et al.}, \emph{Quarkonium mass splittings in three-flavor lattice QCD}, Phys. Rev. D {\bf 81}, 034508 (2010) [{hep-lat/0912.2701}].
\bibitem{lattice_levkova} L.~Levkova and C.~DeTar, \emph{Charm annihilation effects on the hyperfine splitting in charmonium}, Phys. Rev. D {\bf 83}, 074504 (2011) [{hep-lat/1012.1837}].
\bibitem{lattice_kawanai} T.~Kawanai and S.~Sasaki, \emph{Charmonium potential from full lattice QCD}, Phys. Rev. D {\bf 85} 091503(R) (2012) [{hep-lat/1110.0888}].
\bibitem{belle_etacp} S.-K.~Choi {\it et al.} (Belle Collaboration), \emph{Observation of the $\eta_c(2S)$ in 
exclusive $B\rightarrow KK_SK^-\pi^+$ decays}, Phys. Rev. Lett. {\bf 89}, 102001 (2002).
\bibitem{babar_etacp} B.~Aubert {\it et al.} (BaBar Collaboration), \emph{Measurements of the mass and width of the $\eta_c$
meson and of an $\eta_c(2S)$ candidate}, Phys. Rev. Lett. {\bf 92}, 142002 (2004).
\bibitem{cleo_etacp} D.~M.~Asner {\it et al.} (CLEO Collaboration), \emph{Observation of $\eta_c^\prime$ production in $\gamma\gamma$ fusion at CLEO}, Phys. Rev. Lett. {\bf 92}, 142001 (2004).
\bibitem{babar_etacp2} B.~Aubert {\it et al.} (BaBar Collaboration), \emph{Measurement of double charmonium production in $e^+e^-$ annihilations at $\sqrt{s}=10.6$ GeV}, Phys. Rev. D {\bf 72}, 031101 (2005).
\bibitem{belle_etacp2} K.~Abe {\it et al.} (Belle Collaboration), \emph{Observation of double $c\bar{c}$ production in
$e^+e^-$ annihilation at $\sqrt{s}=10.6$~GeV}, Phys. Rev. Lett. {\bf 89}, 142001 (2002).
\bibitem{cb_etacp} C.~Edwards {\it et al.}, \emph{Observation of an $\eta_c^\prime$ candidate with mass 3592$\pm$5~MeV}, Phys. Rev. Lett. {\bf 48}, 70 (1982).
\bibitem{babar_kkpi} B.~Aubert {\it et al.}, \emph{Study of $B$-meson decays to $\eta_c K^{(*)}$, $\eta_c(2S)K^{(*)}$, and $\eta_c\gamma K^{(*)}$}, Phys. Rev. D {\bf 78}, 012006 (2008).
\bibitem{cleo_chicgg} K.~M.~Ecklund {\it et al.} (CLEO Collaboration), \emph{Two-photon widths of the $\chi_{cJ}$ states of charmonium}, Phys. Rev. D {\bf 78}, 091501(R) (2008).
\bibitem{cleo_jpsi_3g} G.~S.~Adams {\it et al.} (CLEO Collaboration), \emph{Observation of $J/\psi\rightarrow 3\gamma$}, Phys. Rev. Lett. {\bf 101}, 101801 (2008).
\bibitem{X3872Belle2003isospin} S.-K. Choi {\it et al.} (Belle Collaboration), \emph{Observation of a narrow charmoniumlike state in exclusive ${B}^\pm\rightarrow K^\pm\pi^{+}\pi^{-}J/\psi$ Decays}, Phys. Rev. Lett. {\bf 91}, 262001 (2003).
\bibitem{Hanhart2009} F.-K.~Guo, C.~Hanhart, and U.-G.~Mei{\ss}ner, \emph{Extraction of the light quark mass ratio from the decays 
$\psi^{'}\rightarrow J/\psi\pi^{0}\eta$}, Phys. Rev. Lett. {\bf 103}, 082003 (2009) [{hep-ph/0907.0521}]; {\it ibid.} {\bf 104}, 109901(E) (2010).
\bibitem{Hanhart2010} F.-K.~Guo {\it et al.}, \emph{Novel analysis of the decays $\psi^{'}\rightarrow h_{c}\pi^{0}$ and 
$\eta_{c}^{'}\rightarrow\chi_{c0}\pi^{0}$}, Phys. Rev. D {\bf 82}, 034025 (2010) [{hep-ph/1002.2712}].
\bibitem{babar_y4260} B.~Aubert {\it et al.} (BaBar collaboration), \emph{Observation of a broad structure in the $\pi^+\pi^- J/\psi$ mass spectrum around 4.26~GeV/$c^2$}, Phys. Rev. Lett. {\bf 95}, 142001 (2005).
\bibitem{babar_y42602} J.~P.~Lees {\it et al.} (Babar Collaboration), \emph{Study of the reaction $e^+e^-\rightarrow J/\psi \pi^+\pi^-$ via initial-state radiation at BABAR}, Phys. Rev. D {\bf 86}, 051102(R) (2012).
\bibitem{cleo_y4260} Q.~He {\it et al.} (CLEO Collaboration), \emph{Confirmation of the Y(4260) resonance production in initial state radiation}, Phys. Rev. D {\bf 74}, 091104(R) (2006).
\bibitem{cleo_y42602} T.~E.~Coan {\it et al.} (CLEO Collaboration), \emph{Charmonium decays of $Y(4260)$, $\psi(4160)$, and $\psi(4040)$}, Phys. Rev. Lett. {\bf 96}, 162003 (2006).
\bibitem{belle_y4260} C.~Z.~Yuan {\it et al.} (Belle Collaboration), \emph{Measurement of the $e^+e^-\rightarrow \pi^+\pi^- J/\psi$ cross section via initial-state radiation at Belle}, Phys. Rev. Lett. {\bf 99}, 182004 (2007).
\bibitem{bes3_x3900} M.~Ablikim {\it et al.} (BESIII Collaboration), \emph{Observation of a charged charmoniumlike structure in 
$e^+e^-\rightarrow \pi^+\pi^- J/\psi$ at $\sqrt{s}=4.26$~GeV}, Phys. Rev. Lett. {\bf 110}, 252001 (2013) [{hep-ex/1303.5949}].
\bibitem{belle_x3900} Z.~Q.~Liu {\it et al.} (Belle Collaboration), \emph{Study of $e^+e^-\rightarrow \pi^+\pi^- J/\psi$ and observation
of a charged charmonium-like state at Belle}, Phys. Rev. Lett. {\bf 110}, 252002 (2013) [{hep-ex/1304.0121}].
\bibitem{cleo_x3900} T.~Xiao, S.~Dobbs, A.~Tomaradze, and Kamal~K.~Seth, \emph{Observation of the charged hadron $Z_c(3900)$ at $\sqrt{s}=4170$~MeV}, [{hep-ex/1304.3036}]. 

\end{thebibliography}
\end{document}